\documentclass[11pt,a4paper]{article}
\usepackage{graphicx,bm}
\begin{document}
\title{Quantum creep in layered antiferromagnetic superconductor}
\author{T. Krzyszto\'n \\
\small {Institute of Low Temperature and Structure Research} \\
\small {Polish Academy of Sciences} \\
\small {50-950 Wroc\l{}aw, P.O.Box 1410,Poland}}
\maketitle
\begin{abstract}In the mixed state of layered superconductor the antiferromagnetic order of magnetic ions can create the spin-flop domains along the phase cores of the Josephson vortices. The paper discusses how this feature affects the macroscopic quantum tunnelling of the Josephson vortices. It is shown that the action and hence the activation energy is rendered temperature dependent so that the quantum tunnelling rate becomes temperature dependent below the crossover temperature. It is also shown that in constant temperature thermal or quantum creep may occur, depending on the direction or intensity of the applied magnetic field.
\end{abstract}

\section*{Introduction}
The discoveries of ternary Rare Earth (RE) Chevrel Phases REMo$_{6}$S$_{8}$ and RERh$_{4}$B$_{4}$ \cite{Ternary,BulBuzdKulPanj} compounds with regular distribution of localized magnetic moments of RE atoms have proved conclusively the coexistence of various types of magnetism with superconductivity. Intensive experimental and theoretical research has shown that 4f electrons of RE atoms responsible for magnetism and 4d electrons of molybdenum chalcogenide or rhodium boride clusters responsible for superconductivity are spatially separated and therefore their interaction is weak. In many of these systems superconductivity coexists rather easily with antiferromagnetic order, where usually the Neel temperature $T_{N}$ is lower than the critical temperature for superconductivity $T_{c}$. For almost two decades the problem of the interaction between magnetism and superconductivity has been overshadowed by high temperature superconductivity (HTS) found in copper oxides. However, the discovery of the magnetic order in Ru-based superconductors \cite{Bauer,Pringle,KlamutX,Houzet} inspired a return to the so-called coexistence phenomenon \cite{Maple95}. The interplay between magnetism and superconductivity was studied in d-electron UGe$_{2}$\cite{Saxena} and ZrZn$_{2}$\cite{Pfleiderer}, where itinerant ferromagnetism may coexist with superconductivity, and in heavy fermion UPd$_{2}$Al$_{3}$\cite{Sato}, where magnetic excitons are present in the superconducting phase. The recent discovery of the iron pnictide superconductors \cite{Kamihara} have triggered broad interest in the mechanism of the coexistence of magnetism and superconductivity in this new class of superconductors \cite{Takeshita}. They exhibit qualitative similarity to cuprates in that superconductivity occurs upon carrier doping (electrons, in this case) of pristine compounds that exhibit magnetism \cite{Nakai}.

Among classic magnetic superconductors, the Chevrel phases have been studied most intensively. These compounds are mainly polycrystalline materials. However, some specific effects can be measured only on single crystals. One such effect is a two-step flux penetration process, predicted in Ref.\cite{Krzy80,Krzy84} and after that discovered in the antiferromagnetic superconductor (bct) ErRh$_{4}$B$_{4}$ \cite{Muto86}. Later it was observed also in DyMo$_{6}$S$_{8}$ \cite{K&R2002}. That specific effect is a consequence of creation of the spin-flop (or metamagnetic) domain along the vortex core.

Consider as an example antiferromagnet with two magnetic sublattices. An infinitesimal magnetic field applied perpendicular to its easy axis makes the ground state unstable against the phase transformation to the canted phase (spin-flop). On the other hand, if the magnetic field is applied parallel to the easy axis the antiferromagnetic configuration is stable up to the thermodynamic critical field $H_{T}$. When the field is further increased a canted phase develops in the system. Assume that in the antiferromagnetic superconductor the lower critical field fulfils the relation $H_{c1}<{\frac12}H_{T}$ and that the external field,$H_{c1}<H<{\frac12}H_{T}$, is applied parallel to the easy axis. Then the superconducting vortices appear in the ground antiferromagnetic state. If the field is increased above ${\frac12}H_{T}$ in the core originates the phase transition to the canted phase because the field intensity in the core doubles the intensity of the external field~\cite{ClemCoffey90}. The spatial distribution of the magnetic field of the vortex is a decreasing function of the distance from the center of the vortex. Hence the magnetic field intensity in the neighborhood of the core is less then $H_{T}$. Therefore, the rest of the vortex remains in the antiferromagnetic configuration. The radius of spin flop domain grows as the external field is increased.

Thus, in the considered model there are two distinct types of vortices. Possible candidate of such system might be $ErBa_{2}Cu_{3}O_{7}$, however, the features described in this paper have not yet been experimentally investigated. That compound has tetragonal unit cell with small orthorombic distortion in the $ab$ plane. The $Er$ ions form two sublattices antiferromagnetic structure of magnetic moments laying parallel and antiparallel to the $\mathbf{a}$ direction \cite{Lynn93}. We shall use in this paper an abstract structure which resembles the above mentioned structure of $ErBa_{2}Cu_{3}O_{7}$. It consists of the superconducting layers of thickness $d_{s}$ and the isolating layers of thickness $d_{i}$, $d=d_{i}+d_{s}$. In the isolating layers, the magnetic moments are running parallel and antiparallel to the $\mathbf{a}$ direction (easy axis). The magnetic field aligned parallel to the conducting planes makes the vortex lattice accommodate itself to the layer structure so that the vortex cores are lying between the superconducting sheets.

\begin{center}
\begin{figure}
\includegraphics*[width=0.8\textwidth]{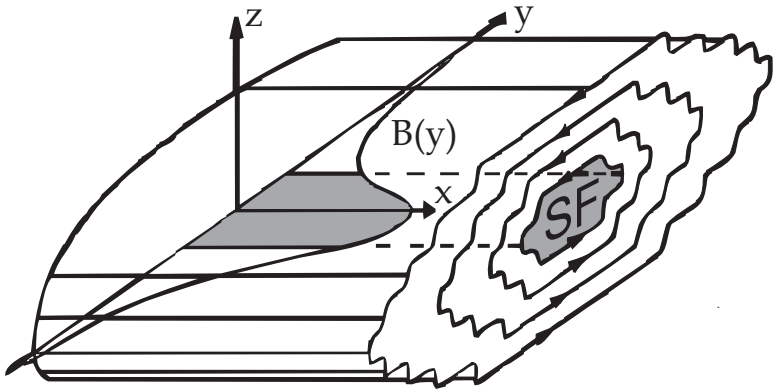}
\caption{Single Josephson vortex lying in the $ab$ plane along the $\hat{x}$-axis ($\hat{a}$-axis). The SF domain induced along the
phase core is shown in the gray area.}
\label{fig1}
\end{figure}
\end{center}

The structure of a vortex lying in the $\mathbf{ab}$ plane in a layered superconductor with Josephson coupling between adjacent layers resembles the Abrikosov`s one except that the order parameter does not vanish anywhere \cite{ClemCoffey90}. Instead there exists a region where the Josephson current $j_{z}$ is of the order of the critical current. In this region, named the phase core, the London model fails as in the classic superconductor. Away from the phase core the streamlines of the shielding supercurrents, which also represents contours of constant magnetic field, are elliptical except for the zigzags, shown in Fig.(~\ref{fig1}), due to the intervening insulating layers.
\section*{Macroscopic quantum tunnelling through intrinsic pinning barriers}
A current density $j$, flowing along the planes perpendicular to the applied magnetic field exerts a Lorentz force on the vortices in the $\mathbf{c}$ direction so that intrinsic pinning barriers are formed on superconducting layers. The experimental evidence of quantum tunnelling is based on the fact that the magnetic moment relaxation rate exhibits two types of behavior as a function of temperature. Above a characteristic temperature $T_{0}$ in the thermal activation regime the decay rate is of the Arrhenius type $\Gamma \sim \exp \left( -U_{0}/k_{B}T\right)$. Below $T_{0}$, the decay rate is assumed {\it a priori}to be essentially  independent of temperature $\Gamma \sim \exp \left( -S/\hbar \right)$ and is interpreted as arising from the quantum tunneling of vortices through intrinsic pinning potential \cite{Blatter94,Ivlev91,Hoekstra99,Barone2010,Morais}. In the following we show a considerable change of tunneling rate and crossover temperature due to the SF phase formation in the vortex core.
We shall assume that the vortex line is a straight string-like object, with effective mass $m$ per unit length, moving in a metastable intrinsic pinning potential $V(u)$ and exposed to continuous deformation $u(x,t)$ in the $\hat{z}$ direction. The magnetic field is applied in $\hat{x}$ direction. Following Caldeira and Leggett \cite{Caldeira83} the vortex is coupled to a heat-bath reservoir of harmonic oscillators interacting linearly with the vortex. In the semi-classical approximation the quantum decay rate is calculated as a saddle-point solution (bounce) of the Euclidean action $S$ for the string
\begin{eqnarray}
S = \int_{-\infty }^{\infty }dx\int_{0}^{\hbar \beta } d\tau\left\{ \frac{1}{2} m\left( \frac{\partial u}{\partial \tau }\right)
^{2}+\frac{\varepsilon _{l}}{2}\left( \frac{\partial u}{\partial x}\right)^{2}+V(u)\right. \nonumber \\
\left. - \frac{\eta }{2\pi }\frac{\partial u}{\partial \tau }\int_{0}^{\hbar \beta }d\tau ^{^{\prime }} \frac{\partial u}{\partial
\tau ^{^{\prime }}}\ln \left| \sin \frac{\pi }{ \hbar \beta }\left( \tau -\tau ^{^{\prime }}\right) \right| \right\}
\label{eq1}
\end{eqnarray}
Here $\beta =\left( k_{B}T\right) ^{-1}$ , $\eta $ is the viscosity coefficient and $\tau $ denotes imaginary time. The pinning potential $V(u)$ consists of intrinsic periodic part and the Lorentz potential:
\begin{equation}
V\left( u\right) =-\frac{\varphi _{0}j_{c}d}{2\pi }\cos \left( \frac{2\pi u}{ d}\right) -\varphi _{0}ju,
\label{eq2}
\end{equation}
where $j_{c}$ denotes critical depinning current. For large current, this potential can be expanded around the inflection point to give
\begin{equation}
V\left( u\right)= V_{0}\left[ \left( \frac{u}{w}\right)^{2}-\left( \frac{u}{w}\right)^{3}\right],
\label{eq3}
\end{equation}
where
$V_{0}=\frac{2}{3}\frac{\varphi_{0}j_{c}^{2}\pi ^{2}}{d^{2}}w^{3}$ and $w=\frac{3d}{\pi }\left(
\frac{j_{c}-j}{2j_{c}}\right)^{\frac{1}{2}}$ may be thought as the height and width of the barrier (because $V(0)=V(w)=0$), and $j_{c}$ is the critical depinning current. The last term in Eq.(\ref{eq1}) is so called Caldeira-Leggett action, which describes ohmic damping produced by the coupling to harmonic oscillators. The line tension $\varepsilon _{l}$ is different for vortices in two different orientations ($\mathbf{a}$ and $\mathbf{b}$)  in the $\mathbf{ab}$ plane. In the semiclassical approximation the decay rate is given by the value of the action on a classical trajectory obtained from the Euler-Lagrange equations of the motion \cite{Grabert}.
\begin{equation}
m\frac{\partial ^{2}u}{\partial \tau ^{2}}+\varepsilon _{l}\frac{\partial
^{2}u}{\partial x^{2}}-V^{^{\prime }}(u)-\frac{\eta }{\hbar \beta}\int_{0}^{\hbar \beta }d\tau \frac{\partial u}{\partial \tau }\cot \frac{\pi }{\hbar \beta }\left( \tau -\tau ^{^{\prime }}\right)=0
\label{eq4}
\end{equation}
 In the thermal regime $T>T_{cr}$ the classical trajectory $u_{0}(x)$ gives the activation energy $U_0$ from the static solution of the following equation
\begin{equation}
-\varepsilon _{l}\frac{\partial ^{2}u_{n}}{\partial x^{2}}+V^{^{\prime}}(u_{0})=0.
\label{eq5}
\end{equation}
Below the crossover temperature $T<T_{cr}$ a new kind of trajectory, periodic in the imaginary time, develops. Therefore, $u(x,\tau )$ can be expanded in the Fourier series with Matsubara frequencies
\begin{equation}
u(x,\tau )=\sum_{n=0}^{\infty }u_{n}\left( x\right) \cos \left( \omega _{n}\tau \right) \ \ \ ; \ \ \omega _{n}=\frac{2\pi n}{\hbar
\beta }.
\label{eq6}
\end{equation}
Substituting this expansion into Eq.(\ref{eq3}) and linearizing potential around the static solution $u_{0}(x)$ one obtains
\begin{equation}
 -\varepsilon _{l}\frac{\partial ^{2}u_{n}}{\partial x^{2}}+V^{^{\prime \prime }}(u_{0})u_{n}=-\left( \eta
 \omega_{n}-m\omega_{n}^{2}\right) u_{n}.
 \label{eq7}
 \end{equation}
 Upon introducing new variables $v_{n}=\frac{u_{n}}{w}$ \\ and $\zeta =\frac{x}{d}\left( \frac{\pi ^{2}w\varphi _{0}j_{c}}{\varepsilon
 _{l}}\right)^{\frac{1}{2}}$ the static equation now reads.
 \begin{equation}
 -\frac{1}{2}\frac{\partial ^{2}v_{0}}{\partial \zeta ^{2}} +2v_{0}-3v_{0}^{2}=0
 \label{eq8}
 \end{equation}
 Its solution is easily found to be
 \begin{equation} v_{0}=\cosh ^{-2}\zeta
\label{eq9}
\end{equation}
Substitution Eq.(6) into Eq.(4) results in the following equation
\begin{equation}
-\frac{1}{2}\frac{\partial ^{2}v_{n}}{\partial \zeta ^{2}}+2\left( 1-3\cosh^{-2}\zeta \right) v_{n}=E_{n}v_{n},
\label{eq10}
\end{equation}
where
\begin{equation}
E_{n}=-\frac{j_{c}w^{2}}{V_{0}^{2}}\left( \eta \omega _{n}+ m\omega_{n}^{2}\right).
\label{eq11}
\end{equation}
Eq.(\ref{eq9}) has three discrete eigenvalues: $-\frac{5}{2},0,\frac{3}{2}$ \cite{Landau}. The negative one determines the crossover temperature
\begin{equation}
k_{B}T_{cr}=\frac{\hbar \eta }{4\pi m}\left(\sqrt{ 1+14.2\frac{\pi \varphi _{0}j_{c}m}{d\eta ^{2}}\sqrt{ 1 - \frac{j}{j_c } }} -1 \right).
\label{eq12}
\end{equation}
In the thermally activated region above $T_{cr}$ we use the static solution of Eq.(\ref{eq5}) and get the action $S_0$ which arises only from the elastic and pinning terms of Eq.(\ref{eq1}). The action $S_0 = U_0\hbar \beta$ is then given by:
\begin{equation}
S_{0} = 2.77 \left( \frac{\varepsilon_{l} d^3
\phi_{0}j_{c}}{\pi^3}\right)^{\frac{1}{2}}\left( {1 - \frac{j}{{j_c }}}
\right)^{5/4}\hbar\beta.
\label{eq13}
\end{equation}
Below $T_{cr}$, the predominant mechanism of the decay of the meta-stable state is the quantum-mechanical tunneling, and the action on the bounce trajectory determines the rate of this process. To include contributions of effective mass and viscous damping in the bounce action it is necessary to go beyond the harmonic approximation in Eq.(\ref{eq7}) and substitute for $u(x,\tau )$ the leading terms in the Matsubara frequencies. Calculations are performed in the vicinity of crossover temperature so the perturbation parameter is $\left(1-T/T_{cr}\right)$. It can be shown that $v_{n}$ are of the order $\left( 1-T/T_{cr}\right)^{\frac{n}{2}}$ and the leading-order contribution comes from $v_1\sim cosh^{-3}\zeta$. Substituting this term for $u$ in Eq.(\ref{eq1}) one can obtain the total action for the bounce trajectory involving contribution of the inertial mass and the viscous damping.
\begin{eqnarray}
S &=& 5.26\hbar \beta \left[{\frac{{\varepsilon _l d^3\varphi _0 j_c }}{\pi }} \right]^{1/2} \left( {1 - \frac{j}{{j_c }}}
\right)^{5/4}\nonumber \\
 &+&  \left[ {\frac{{\varepsilon _l d^5}}{{\pi \varphi _0 j_c }}} \right]^{1/2} \left( {5.71\frac{m}{\hbar \beta}
 + 2.85\frac{\eta }{\pi }} \right)\left( {1 - \frac{j}{{j_c }}} \right)^{3/4}
 \label{eq14}
\end{eqnarray}
\section*{Evaluation of the parameters}
The above calculations apply to both kind of vortices characterized by different effective masses, line tensions and viscosity coefficients. We shall calculate these parameters as functions of condensation energy accumulated in the vortex cores. For the stationary flux flow the viscous force $ \eta \frac{\partial u}{\partial t}$ is equal to Lorentz force. The electric field generated by the moving vortex is $E=B\frac{\partial u}{\partial t}$, so we get $E=\frac{\varphi _{0}B}{\eta }j=\rho j=\rho _{N}\frac{B}{H_{c2}}j$ where $\rho _{N}$ is the normal phase resistivity in the $\mathbf{ab}$ plane and $ H_{c2}$ is the upper critical field parallel to the layers. Finally,
\begin{equation}
\eta =\frac{\varphi _{0}H_{c2}}{\rho _{N}}=\frac{\varphi _{0}\kappa H_{c} \sqrt{2}}{\rho _{N}}=\varepsilon _{l}\frac{4\sqrt{3}\kappa
^{2}}{\pi \rho _{N}\ln \kappa },
\label{eq15}
\end{equation}
where $H_{c}=\frac{\varepsilon _{l}\kappa 2\sqrt{6}}{\pi \varphi _{0}\ln \kappa }$ is calculated from the constitutive relation $\varepsilon _{l}=H_{c1}\varphi _{0}$.

A moving vortex in the magnetic superconductor can transfer energy to the magnetic moments by emitting spin waves. This energy transfer gives rise to magnetic contribution to the mass and viscosity of the vortex \cite{Bulaevskii2010}. However the vortex velocity for this effect to occur should exceed the velocity of the spin wave which is not the case during the flux creep.

The effective mass of the vortex can be deduced from the work of Suhl\cite{Suhl}. He derived the core contribution to the inertial mass $m=\frac{3}{8}m_{e}\frac{\xi ^{2}H_{c}^{2}\mu _{0}}{ \epsilon _{F}}$, where $m_{e}$ denotes the mass of the electron and $ \epsilon _{F}$ the Fermi energy, and the electromagnetic contribution coming from the energy of the electric field induced by the moving flux. Simple estimation shows that this contribution in layered superconductors is $10^{-4}$ of the core contribution. Therefore,
\begin{equation}
m=\varepsilon _{l}^{2}\frac{9\lambda _{ab}^{2}m_{e}\mu _{0}}{\varphi _{0}^{2}\pi ^{2}\epsilon _{F}\left( \ln \kappa \right) ^{2}}.
\label{eq16}
\end{equation}
The vortices in two main orientations in the $\mathbf{ab}$ plane have different line tensions. These lying parallel to $\mathbf{b}$ direction and laying in the $\mathbf{a}$ direction, but created in the magnetic field less then $\frac{1}{2}H_{T}$, have the line tension $\varepsilon _{b}$,  and the ones lying in the $\mathbf{a}$ direction but possessing SF domain have the line tension $\varepsilon _{a}$.
\begin{equation}
\varepsilon _{b}=\epsilon _{0}\ln \frac{\lambda_{ab}}{d}; \\ \varepsilon _{a}=\frac{\varphi_{0}H_{T}}{2}+\frac{9}{128}\epsilon _{0}\ln \frac{\varphi _{0}\lambda_{c}^2}{\pi (\mu_0 H_{T}+M)d^2\lambda_{ab}^2},
\label{eq17}
\end{equation}
where $\epsilon_{0}=\frac{\varphi_{0}^2}{4\pi\lambda_{ab}\lambda_{c} \mu_{0}}$, $M$ is the magnetic moment of the SF domain inside the vortex and $H_{T}$ is the thermodynamic critical field for spin-flop transformation. In order to evaluate the above line tensions it is necessary to calculate $H_{T}$ and $M$. Therefore a following argumentation is proposed. At low fields, in the vicinity of the lower critical field $H_{c1}$, the intensity of the field in the vortex core is $2H_{c1}$~\cite{ClemCoffey90}. When the external field is increased, the intensity of the magnetic field in the vortex core increases due to the superposition of the fields of the surrounding vortices. The field intensity in the core must reach $H_T$ in order to originate a transition to the SF phase. Thus, taking into account only $z$ nearest neighbors we can write for the nonunilateral triangular lattice
\begin{eqnarray}
\varphi_{0}H_{T}&=&2\varphi_{0}H_{c1}+4z\varepsilon_b\left(\ln\frac{\lambda_{ab}}{d}\right)^{-1} \nonumber \\
&\times&\left[K_{0}\left(\frac{c}{\lambda_{ab}}
\right)+2K_{0}\left(\frac{c}{2\lambda_{ab}}\sqrt{\frac{3\lambda_c}{\lambda_{ab}}}\right)\right] \nonumber \\
&=&2\varepsilon_b + o\left(\varepsilon_b\right),
\label{eq18}
\end{eqnarray}
where $c$ is the lattice constant. Although there are no precise measurements of the spin-flop transition in the antiferromagnetic high temperature superconductors, we assume that $ \mu _0H_T\approx40mT $. The typical value of $ 4.2\mu _B$ per $Er$ ion per unit cell in $ErBa_{2}Cu_{3}O_{7}$ \cite{Abulafia} gives $ M \approx 0,35~\rm{T}$ and $B_T\approx 0,40~\rm{T}$. Since $\varphi_0 H_{c1}=\varepsilon_b$ and $d/\xi_c \approx 1$ we obtain:
\begin{equation}
\frac{\varepsilon_a}{\varepsilon_b}=1+\frac{36}{128}\frac{\ln \left( \frac{\displaystyle{\varphi_{0}\lambda_{c}^2}} {\displaystyle{\pi(\mu_0 H_{T}+M)d^2\lambda_{ab}^2}} \right)}{\ln\left(\frac{\displaystyle{\lambda_{ab}}}{\displaystyle{d}}\right)} \approx1.7 \label{eq19}
\end{equation}
It is possible now to relate the viscosity coefficient Eq.(\ref{eq15}) and the mass of the vortex Eq.(\ref{eq16}) to its line tension Eq.(\ref{eq17}).
\begin{equation}
\frac{\eta_a}{\eta_b}=\frac{\varepsilon_a}{\varepsilon_b}
\label{eq20}
\end{equation}
and
\begin{equation}
\frac{m_{a}}{m_{b}}=\left( \frac{\varepsilon_a}{\varepsilon_b}\right) ^2 ,
\label{eq21}
\end{equation}
With the use of a simple consideration we can estimate the change of $j_c\ $ due to the creation of spin-flop domain along the vortex
\[ j_{c}\varphi _{0}d\sim \frac{1}{2}\int dydzC(y,z)\left( \frac{\partial u_{z}}{\partial z}\right) ^{2}, \]
where the Fourier transform of the compression modulus is given by \cite{Brandt}
\[
C(k_{y},k_{z})=\frac{B^{2}}{\mu _{0}(1+\lambda _{ab}^{2}k_{z}^{2}+\lambda
_{c}^{2}k_{y}^{2})}\]
By taking $dydz\sim \varphi _{0}/B ; u_{z}\sim d ; \frac{\partial }{\partial z} \sim k_{z}\sim k_{y}(\lambda _{c}/\lambda _{ab})$ the estimation of the integral gives
\begin{equation}
j_{c}=\frac{Bd}{4\lambda _{ab}^{2}}.
\label{eq22}
\end{equation}
In the $\mathbf{a}$ direction, however, we have an additional contribution from the magnetic domain $dydz\sim
5\varphi _{0}/8 \pi B_{T}$ , so we get
\begin{equation}
j_{ca}=j_{c}+\frac{5dB_{T}}{128\lambda _{ab}^{2}}
\label{eq23}
\end{equation}
and finally $ j_{ca}\approx 3j_{cb}$.
\section*{Motion of the flux in the quantum regime}
Consider a hollow cylindrical sample of the radius $a$,  and wall thickness $g << a$ (thin wall approximation). The sample is placed in a magnetic field directed parallel to the axis of the cylinder and along the superconducting layers. A trapped flux in the system is $ \Phi \cong (B_{in}-B_{ex})$, where $B_{in}$ denotes the field inside the hole of the sample and $B_{ex}$ outside the sample respectively. The motion of the flux is triggered off by an activation process in which segments of the flux line tunnel through an intrinsic pinning potential to the neighboring interlayer spacing. By applying Faraday's law we can easily calculate the electric field in the sample due to the change of the trapped flux
\[
E=-\frac{1}{2}\mu_0 ag j_c \frac {d (j/j_c)}{dt}
\]
which is equal to the mean electric field associated with the motion of vortices $E= \varphi_0 W L g$, where $L$ is the length of the sample in the direction of the applied field, and $W$ is the activation probability per unit volume and unit time. As is shown in \cite{Simanek} activation probability in the weak damping approximation is equal to $ \sqrt{\frac {30 S_0}{\pi \hbar}}\exp \left(- \frac{S}{\hbar}\right) $. Combining all above together we get
\begin{equation}
\Omega t = -\int_{x(0)}^{x(t)} \frac{\exp \left(S/\hbar \right)}{\sqrt {S_0/\hbar}} dx
\label{eq24}
\end{equation}
where $x=j/j_c$ ; $\Omega = 38.83 (\frac{\varphi _{0}}{\mu _{0}j_c}) (\frac{L}{2\pi a})$, and $2\pi a$ is the length of the sample along the flow of the current.
\section*{Numerical results}
The purpose of the present study is to derive expressions for the bounce contribution to the action arising from the inertial mass term, elastic and pinning terms, and the damping term. As was mentioned previously, in antiferromagnetic superconductors may occur vortices of two type. One type quite similar to the vortices in nonmagnetic superconductor and the other one possessing a ferromagnetic-like domain around and inside the core. What distinguishes these vortices in the present calculations is their line tension Eq.(\ref{eq19}) and critical depinning current Eq.(\ref{eq23}). The input parameters for numerical calculation are shown in Tab.(\ref{tab1}).

\begin{table} [!htb]
\caption{Input parameters \cite {Hoekstra99}, \cite{Ivlev91}, \cite{Harshman}, \cite{Matsuda}, \cite{Sheng}.}
\begin{center}
\begin{tabular}{*{20}c}
\hline
\hline
   $d = 10^{ - 9} m$ & $\eta_b  = 10^{ - 8} kg/( s \cdot m) $  \\
   $j_{cb}  = 10^{11} A/m^2 $ & $\lambda _{ab}  = 160 nm $ \\
   $m_{b}  = 10^{- 22} kg/m$ & $\lambda _c  = 800 nm $  \\
   $\epsilon _F  = 0.1eV$ & $\mu _0 H_{c2}^{ab}  = 150T$ \\
\hline
\hline
\end{tabular}
\end{center}
\label{tab1}
\end{table}

We consider the limit of a large current, slightly less then critical depining current, and assume $\frac{L}{2\pi a}\sim 1$. In this approximation we calculated the crossover temperature from Eq.(\ref{eq12}) and  plotted it as a function of (FC), fractional current $j/j_c $.
\begin{center}
\begin{figure}[!htb]
\includegraphics*[width=0.8\textwidth]{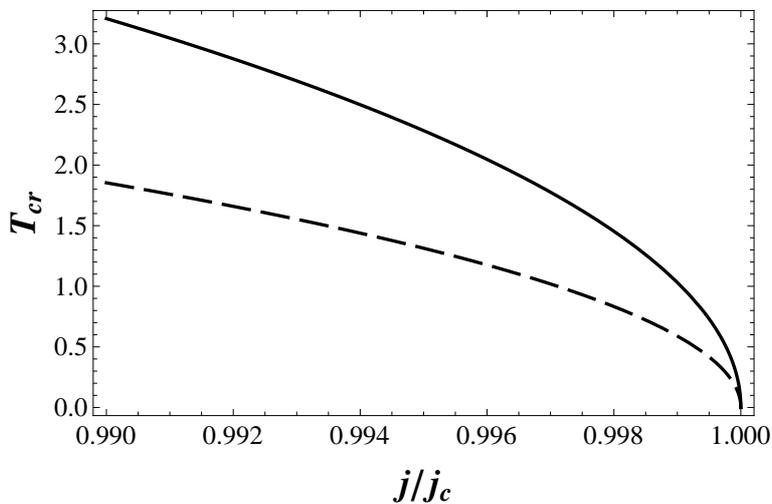}
\caption{Crossover temperature as a function of fractional current $j/j_c$. Dashed line depicts vortices possessing magnetic cores.}
\label{fig2}
\end{figure}
\end{center}
As can be seen the difference of crossover temperatures for both types of vortices depends on the current, and for the value of FC equal to $0.99$  this difference is about $1.3$ K, and vanishes when FC reaches $1$.
\begin{center}
\begin{figure}[!htb]
\includegraphics*[width=0.8\textwidth]{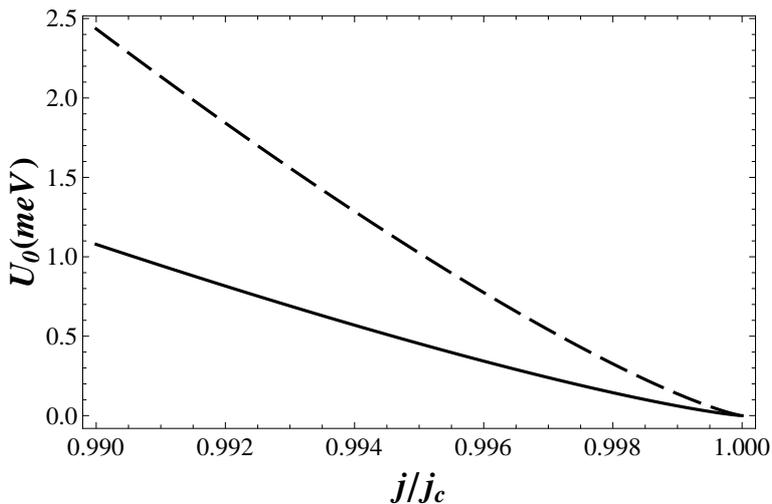}
\caption{Activation energy $U_0$ (in meV) calculated as a function of fractional current $j/j_c $. Dashed line depicts vortices possessing magnetic cores.}
\label{fig3}
\end{figure}
\end{center}
Fig.(\ref{fig3})shows the activation energy $U_0$ calculated from Eq.(\ref{eq13}) as a function of FC. The activation energy differs about $1.5$ meV for $0.99$ of FC and goes to $0$ when FC reaches $1$.

\begin{center}
\begin{figure}
\includegraphics*[width=0.8\textwidth]{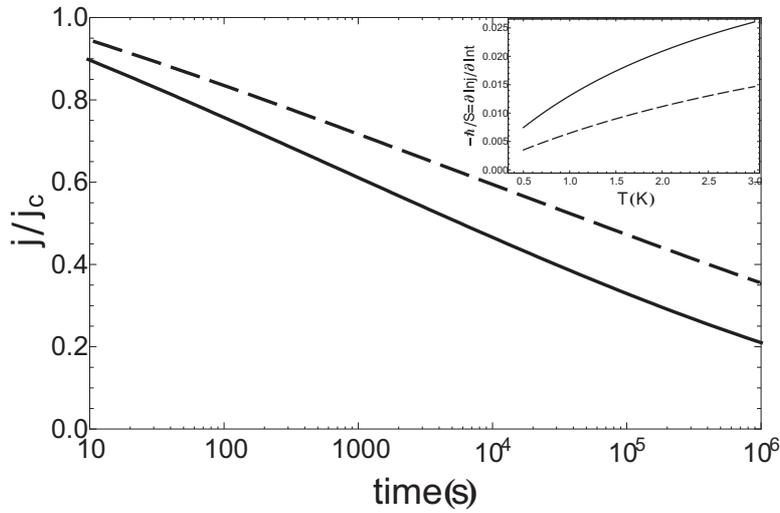}
\caption{Decay of the flux ($\Phi/\Phi_0\sim j/j_c $) as a function of time(logarithmic scale), at fixed temperature 0.5K. In the inset, quantum creep rate as a function of temperature. Dashed line depicts vortices possessing magnetic cores.}
\label{fig4}
\end{figure}
\end{center}

Equation (\ref{eq24}) was solved numerically for the input parameters shown in table (\ref{tab1})
which give the results shown in the Fig.(\ref{fig4}).
The results are consistent with the experimental findings for the YBCO class superconductors \cite{Sheng}. As we can see, the creep is slower when vortices possessing magnetic domain occurred in the system. Another important effect is that the action and hence activation energy is rendered temperature dependent so that the quantum tunneling rate becomes temperature dependent below the crossover temperature.
\section*{The "creep valve" mechanism}
Another conclusion follows from the above considerations. The creep regime can be altered at fixed temperature. It is possible that the system leaps over from quantum to thermal creep regime, or vice versa, when the direction or intensity of the applied magnetic field is changed.
\begin{center}
\begin{figure}[!h]
\includegraphics*[width=0.8\textwidth]{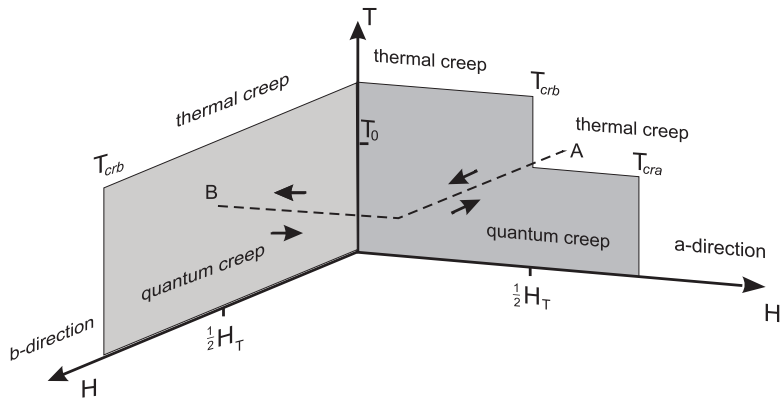}
\caption{Schematic diagram of the leap from thermal to quantum creep, and vice versa.}
\label{fig5}
\end{figure}
\end{center}
To see how it can happen let us fix the temperature of the sample somewhere in the range $T_{crb}>T_{0}>T_{cra}$, as shown in Fig.(\ref{fig5}), and increase magnetic field intensity in the $\mathbf{a}$ direction to the point marked A. Now the system is in thermal creep regime. Then we change the direction of external field from $\mathbf{a}$ to $\mathbf{b}$ direction. The system now leaps over to the point B and finds itself in the quantum creep regime. Doing the same operation in the reverse order one enforces the system to crossover from quantum to thermal creep. Another scenario is also possible. When magnetic field is applied along $\mathbf{a}$ direction and the temperature is fixed in the interval $T_{crb}>T_{0}>T_{cra}$, the increase of magnetic field intensity above ${\frac12}H_{T}$ enforces the appearance of magnetic structure in vortices, and changes the quantum activation to thermal one. Lowering the field intensity one can change back the creep from thermal to quantum regime. The described scenario could be named a "creep valve" because creep rate can be decreased or increased by means of sweeping the field around the ${\frac12}H_{T}$ value.
\section*{Summary}
We discussed quantum tunneling of vortices in layered high temperature superconductor. When the damping and inertial mass of the vortex are considered, the calculation shows that the activation energy is rendered temperature dependent so that the quantum tunnelling rate becomes temperature dependent below the crossover temperature. In the mixed state of layered superconductor the antiferromagnetic order of magnetic ions can create the spin-flop domains along the phase cores of the Josephson vortices, and this effect makes an impact on the creep rate in superconductor. The activation of the creep at fixed temperature can be either thermal or quantum depending on the intensity of the applied magnetic field.

\end{document}